\documentclass[useAMS,usenatbib]{mnras}
\usepackage{graphicx}
\usepackage{dcolumn}
\usepackage{bm}
\usepackage[usenames]{color}

\usepackage{amsmath}
\usepackage{amssymb}

\DeclareMathOperator{\arcosh}{arcosh}

\title[The central region of a void: analytical solution]{The central region of a void: an analytical solution}

\author[A. N. Baushev]{A. N. Baushev$^{1}$\\
    $^{1}$Bogoliubov Laboratory of Theoretical Physics, Joint Institute for Nuclear Research,
    141980 Dubna, Moscow Region, Russia}
\begin{document}

\date{}

\pagerange{\pageref{firstpage}--\pageref{lastpage}} \pubyear{2021}

\maketitle

\label{firstpage}

\begin{abstract}
We offer an exact analytical equation for the void central region. We show that the central density
is solely determined by the amplitude of the initial perturbation. Our results suggest that N-body
simulations somewhat overestimate the emptiness of voids: the majority of them should have the
central underdensity $\delta_c > -73\%$, and there should be almost no voids with $\delta_c <
-88\%$. The central region of a void is a part of an open Friedmann's 'universe', and its evolution
differs drastically from the Universe evolution: there is a long stage when the curvature term
dominates, which prevents the formation of galaxy clusters and massive galaxies inside voids. The
density profile in the void center should be very flat. We discuss some void models obtained by
N-body simulations and offer some ways to improve them. We also show that the dark energy makes the
voids less underdense.
\end{abstract}


\section{Introduction}
Astronomical observations suggest that the Universe at large distances has a honeycomb-like
structure: the galaxies are mainly situated on its walls, while the space inside the 'honeycombs'
seems to be almost empty. These voids are vast roundish areas with the size of $\sim 10-100$~{Mpc}
containing only a few visible galaxies. For instance, less than a hundred galaxies have been
observed \citep{bootesvoid} in one of the largest known voids, the Bo\" otes one, while the
diameter of the object exceeds $100$~{Mpc}. However, the actual matter content of the voids is not
clear, and there can be a lot of matter, which is invisible or almost invisible for the observers.
First, the voids may contain a lot of dwarf or dim galaxies and some rarefied gas. Second, there
can be a large amount of dark matter (hereafter DM).

Analytical investigation of voids has a long history \citep{bertschinger1985, blumenthal1992,
weygaert1993, weygaert2004}. Typically, these works imply the spherical symmetry of the initial
perturbation. Now the void content is usually considered with the help of N-body simulations:
either by void searching in realistic cosmological simulations, or by consideration the evolution
of the future void in the linear approximation, and then the nonlinear evolution is simulated with
N-body codes \citep{goldberg2004}. This method allows to consider arbitrary initial perturbations
without any symmetry implied; however, N-body simulations may suffer from significant numerical
effects \citep{17, 20, 21}, and they do not provide simple analytical relationships describing
voids. The aim of this letter is to show that the average density of the void center may be found
analytically in the very general case.

The metric of a Friedmann universe may be written as $ds^2=c^2dt^2-a^2(t) dl^2$, where $a$ is the
scale factor of the universe and $dl$ is an element of three-dimensional length. We denote all the
quantities related to the present-day Universe by subscript '$0$'. We accept the present-day Hubble
constant $H_0=73$~{km/(s Mpc)}, which corresponds to the critical density
$\rho_{c,0}=\dfrac{3H_0^2}{8\pi G}$. We denote the densities of the matter, radiation, dark energy,
and curvature components of the Universe by $\rho_{M}$, $\rho_{\gamma}$, $\rho_{\Lambda}$,
$\rho_{a}$, respectively\footnote{The curvature density is $\frac{8\pi}{3 c^2} G \rho_{a} =
\dfrac{q}{a^2}$, where $q=-1, 1, 0$ if the universe density is higher, smaller, or equal to the
critical one, respectively.}. It is convenient to use the ratios of these quantities to the
critical density $\rho_{c}$:  $\Omega_{M}\equiv \rho_{M}/\rho_{c}$ etc. The present-day values of
the ratios are estimated from observations $\Omega_{M,0}=\rho_{M,0}/\rho_{c,0}\simeq 0.306$,
$\Omega_{\Lambda,0}\simeq 0.694$ \citep{pdg18}. The Friedmann equation in a general case  may be
written as (see e.g. \citep[eqn. 4.1]{gorbrub1}):
\begin{eqnarray}
H^2(t)=\dfrac{H^2_0}{\rho_{c,0}}\left[\rho_{\Lambda,0}+\rho_{a,0}\left(\dfrac{a_0}{a}\right)^2\!\!+\rho_{M,0}\left(\dfrac{a_0}{a}\right)^3
\right]. \label{26a1}
\end{eqnarray}
The scale factor $a$ is, generally speaking, equal to the radius of curvature of the
three-dimensional space, and therefore is uniquely determined. The only exception is the case when
the three-dimensional space is flat ($\rho_{a} =0$), and the radius of curvature is infinite. Then
we have additional freedom in choosing the value of $a$ arbitrary at one moment of time
\citep{gravitation}. This freedom will be helpful for us.

We assume the validity of the standard $\Lambda$CDM cosmological model: the three-dimensional space
is flat ($\rho_{a} =0$), the dark energy behaves as a non-zero cosmological constant (i.e.,
$p_\Lambda=-\rho_{\Lambda}=\it{const}$). Now $\Omega_{\gamma,0}\sim 10^{-4}$, and we neglect the
radiation term ($\rho_{\gamma} =0$). Then $\Omega_{\Lambda,0}+\Omega_{M,0}=1$. The applicability of
the radiation neglect may seem questionable ($\Omega_{\gamma}$ was much larger in the early
Universe) and will be better substantiated below. Neglecting $\rho_{a}$, we obtain from
(\ref{26a1})
\begin{equation}
H(t)=\frac{da}{a dt} = H_0 \sqrt{\Omega_{\Lambda,0}+\Omega_{M,0}\left(\frac{a_0}{a}\right)^3}.
\label{26a2}
\end{equation}
Integrating this equation, we find the age of the Universe:
\begin{equation}
t_0= \frac{2 H_0^{-1}}{3\sqrt{1-\Omega_{M,0}}}\arcosh\left(1/\sqrt{\Omega_{M,0}}\right)
\label{26a3}
\end{equation}

\section{Calculations}
As all large-scale universe structures, the voids evolve from small adiabatic perturbations
existing in the early Universe. We may use the redshift $z+1=a_0/a(t)$ as a variable, instead of
$t$ or $a(t)$. Let us choose a time moment $z_1$ deeply at the matter-dominated phase of the
Universe ($10<z_1<100$). The perturbation that later forms the void is deeply in the linear regime
at $z_1$, as well as we may neglect the influence of the dark energy and radiation at $z=z_1$. We
denote all the quantities related to $z_1$ by subscript '$1$'.

Apparently, the future void contains a lot of smaller structures, but we will ignore them and
consider the smoothed, averaged density profile of the underdensity. We may do so because the only
component of our system with significant pressure is the dark energy, which is perfectly uniform,
and the small-scale matter substructures do not affect the large-scale gravitational field
\citep{zn2}.

We assume in our derivation that the ratio between DM and baryonic matter remains the same at the
void scale. It is not always true: at $z\sim 1000$, when barionic matter decouples from radiation,
the DM perturbations are much deeper than the baryonic ones, and gas flows down to the potential
wells formed by dark matter. However, we set $z_1\ll 1000$, and the difference in the perturbation
depths should be much smaller at that epoch. Moreover, the DM constitutes $>80\%$ of all matter
content of the Universe, and the error introduced by the supposition of the same behavior of dark
and baryonic matter cannot be very significant, while it essentially simplifies the calculations.

The future void at $z_1$ is just a shallow and extended underdensity. We may describe it by the
characteristic wavelength $\lambda(t)=\lambda_0 a(t)/a_0$. The voids are formed from very extensive
waves  ($\lambda_0 > 20$~{Mpc}), and their lengths have always exceeded by orders of magnitude the
current Jeans length in the Universe \citep{gorbrub2}.

We put the coordinate origin in the point of minimum density inside it and denote the negative
contrast $\vartheta\equiv (\langle\rho_M\rangle - \rho_M)/\langle\rho_M\rangle$ of matter density
$\rho_M$ at this point at $z=z_1$ by $\vartheta_1$; $\vartheta>0$ corresponds to an underdensity
regions. An adiabatic wave at $z=z_1$ with the amplitude $\vartheta_1$ looks like
\begin{equation}
\frac{\Delta\rho_M}{\rho_M}=-\vartheta_1\cos\left(\frac{2\pi x}{\lambda_1}\right)\quad  \Delta
v=\vartheta_1\frac{\lambda_1 H_1}{2\pi}\sin\left(\frac{2\pi x}{\lambda_1}\right)\label{26c1}
\end{equation}
A very general property of the perturbations is that the density variations are phase-shifted to
$\pi/2$ with respect to the velocity variations, and so $\Delta v\simeq 0$ near the maxima and
minima of density.

On all the matter-dominated stage of the Universe evolution the small perturbations grow linearly
$\vartheta\propto a(t)$, and so multiplication $\varpi\equiv |\vartheta| \frac{a_0}{a}= (z+1)
|\vartheta|$ remains constant. Quantity $\varpi$ has simple physical sense. Let us consider an
overdensity of the same wave length $\lambda$ and the same absolute density contrast
$|\vartheta|\equiv |\langle\rho_M\rangle - \rho_M|/\langle\rho_M\rangle$ as the underdensity under
consideration. The density contrast of the overdensity also grows as $\vartheta\propto a(t)$ in the
matter-dominated regime, and we may also introduce $\varpi\equiv (z+1) |\vartheta|$, but, contrary
to the underdensity, the overdensity forms an astronomical object when $\vartheta\simeq 1$. The
redshift $z_{form}$ of the collapse is defined by relation $\vartheta= \varpi/(z_{form}+1)\simeq
1$, and so $z_{form}\simeq\varpi-1$. However, this relationship may be made more accurate:
$\vartheta= \varpi\cdot g(z)/(z+1)$ for $z\le 1$. Here $g(z)$ is a multiplier taking into account
the perturbation growth suppression by the dark energy (see \citep[section 4.4]{gorbrub2});
$g(z)\simeq 1$ if $z>1$ and monotonously decreases with $z$ diminution reaching the value
$g(z=0)\equiv g_0= 0.78$ for $\Omega_{M,0}= 0.306$. We obtain the relationship between $\varpi$ and
$z_{form}$:
\begin{equation}
\varpi= (z_{form}+1)/g(z_{form}). \label{26b2}
\end{equation}
A simplified variant of this equation is $z_{form}=\varpi-1$. The voids are formed from very
extensive waves, which would correspond to structures of mass $10^{14}-10^{16} M_\odot$ if they
were overdensities. These are the masses of large galaxy clusters, and they form at $z_{form}\le
1$. Thus, we may expect that $\varpi\sim 1$ for real voids.

Let us encircle the coordinate origin (i.e, the point of minimum density $\vartheta = \vartheta_1$)
by a sphere of comoving radius $R_1\ll\lambda_1$ (below we show that our consideration is actually
valid until $R_1\simeq\lambda_1/12$). From~(\ref{26c1}) we have $\Delta v\simeq \vartheta_1 H_1 x$,
i.e., $\Delta v$ is proportional to the distance from the coordinate origin and does not depend on
$\lambda$. The second property is especially important: the matter distribution near the point of
minimum density cannot be plane wave~(\ref{26c1}), it should look like
$\Delta\rho_M/\rho_M=-\vartheta\cos(2\pi x/\lambda_x) \cos(2\pi y/\lambda_y) \cos(2\pi
z/\lambda_z)$, where $\lambda_x, \lambda_y, \lambda_z$ are, generally speaking, different. However,
we may see that the velocity field and matter distribution are to first order spherically-symmetric
and do not depend on the underdensity shape. Thus, without loss of generality, we may consider the
spherically-symmetric case $\lambda_x=\lambda_y=\lambda_z$. Then we may write out the second terms
in~(\ref{26c1}):
\begin{equation}
\frac{\overline {\Delta\rho_M}}{\rho_{M,1}}=-\vartheta_1\left(1-\frac{6\pi^2
R_1^2}{5\lambda^2_1}\right);\quad \frac{\Delta v}{R_1}=\vartheta_1 H_1\left(1-\frac{2\pi^2
R_1^2}{3\lambda^2_1}\right)\label{26c2}
\end{equation}
Contrary to the first ones, the second terms depend on $\lambda$, i.e., on the underdensity shape.

Thus, the system is spherically-symmetric, and the matter inside $R_1$ has almost constant density
$(1-\vartheta_1) \rho_{M,1}$, since the sphere surrounds an extremum point of $\rho_M$. If we
neglect the tidal influence of the matter outside $R_1$, we may use the standard Tolman approach
\citep{tolman1934, bondi1947} to describe the universe evolution inside $R_1$, which means that
this area can be considered as a part of a 'universe' with cosmological parameters different from
the ones of the undisturbed Universe. Equation (\ref{26a1}) for this 'universe' looks like
\begin{eqnarray}
\mathcal{H}^2(t)
=\dfrac{H^2_0}{\rho_{c,0}}\left[\gamma_{\Lambda,1}+\gamma_{a,1}\left(\dfrac{b_1}{b}\right)^2\!\!+
\gamma_{M,1}\left(\dfrac{b_1}{b}\right)^3\right]. \label{26a4}
\end{eqnarray}
Here we use notations $\gamma$, $b$, and $\mathcal{H}$ instead of $\rho$, $a$, and $H$
in~(\ref{26a1}), respectively. Contrary to the unperturbed Universe case, $\gamma_{a,1}$ is
essentially not zero.

On the one hand, the speed of the sphere $R_1$, which confines the 'universe', is equal to
$dR/dt=\mathcal{H}(z_1) R_1$ at $z=z_1$. On the other hand, $dR/dt=H(z_1) R_1 +\Delta v(R_1)$,
i.e., the speed of the sphere is the sum of the Hubble expansion of the unperturbed Universe and
the perturbation speed. Thus, $\mathcal{H}(z_1)=H(z_1) +\Delta v(R_1)/R_1$, and substituting here
the first term for $\Delta v(R_1)$ from~(\ref{26c2}), we obtain
$\mathcal{H}^2(z_1)=H^2(z_1)(1+\vartheta_1)^2\simeq H^2(z_1)(1+2\vartheta_1)$, since
$\vartheta_1\ll 1$. Substituting here equations~(\ref{26a2}) and~(\ref{26a4}) at the moment
$z=z_1$, we obtain:
\begin{eqnarray}
\dfrac{\gamma_{\Lambda,1}}{\rho_{c,0}}+\dfrac{\gamma_{a,1}}{\rho_{c,0}}+
\dfrac{\gamma_{M,1}}{\rho_{c,0}}=(1+2\vartheta_1)\left(\Omega_{\Lambda,0}+\Omega_{M,0}\left(\frac{a_0}{a_1}\right)^3\right).
\label{26a5}
\end{eqnarray}
Apparently, $\gamma_{\Lambda,1}=\rho_{\Lambda,1}=\gamma_{\Lambda,0}=\Omega_{\Lambda,0} \rho_{c,0}$.
The matter density $\gamma_{M,1}=(1-\vartheta_1)\rho_{M,1}=(1-\vartheta_1) (a_0/a_1)^3 \Omega_{M,0}
\rho_{c,0}$. Substituting these values to (\ref{26a5}) and neglecting $\vartheta_1
\rho_{\Lambda,1}$ in comparison with much larger $\vartheta_1 \rho_{M,1}$, we find:
\begin{eqnarray}
\gamma_{a,1}= 3 \vartheta_1 \;\Omega_{M,0}\; \rho_{c,0}\left(\frac{a_0}{a_1}\right)^3\!\!= 3
\varpi\; \Omega_{M,0}\; \rho_{c,0}\left(\frac{a_0}{a_1}\right)^2 \label{26a6}
\end{eqnarray}
As we have already mentioned, since the three-dimensional space of the Universe is flat, we have
additional freedom in choosing the value of $a$ arbitrary at one moment of time. We choose
$a_1=b_1$, i.e., $a(z_1)=b(z_1)$. Then we may simplify:
\begin{eqnarray}
 \dfrac{\gamma_{a,1}}{\rho_{c,0}}\left(\dfrac{b_1}{b}\right)^2=3\varpi\Omega_{M,0}\left(\frac{a_0}{b}\right)^2 \label{26a7}\\
 \dfrac{\gamma_{M,1}}{\rho_{c,0}}\left(\dfrac{b_1}{b}\right)^3=(1-\vartheta_1)\Omega_{M,0}\left(\frac{a_0}{b}\right)^3\simeq
 \Omega_{M,0}\left(\frac{a_0}{b}\right)^3\label{26a8}
\end{eqnarray}
We substitute these values to (\ref{26a4}) and obtain the Friedmann equation for the central area
of the void:
\begin{eqnarray}
\mathcal{H}(t) =H_0 \sqrt{\Omega_{\Lambda,0}+3\varpi\Omega_{M,0}\left(\frac{a_0}{b}\right)^2 +
\Omega_{M,0}\left(\frac{a_0}{b}\right)^3}. \label{26a9}
\end{eqnarray}
 It may seem strange that we keep the linear in $\vartheta_1$ term in (\ref{26a6}), while we
neglect it in (\ref{26a8}) and approximate $(1-\vartheta_1)\simeq 1$. This is warranted by the fact
that $\gamma_{a}\propto b^{-2}$, while $\gamma_{M}\propto b^{-3}$. Yes, at $z=z_1$ the terms in
$\gamma_{M}$ and $\gamma_{a}$ linear in $\vartheta_1$ are comparable (and small). However, the
linear term in $\gamma_{M}$ scales as $b^{-3}$ and remains only a small correction to $\gamma_{M}$,
while the linear term in $\gamma_{a}$ scales as $b^{-2}$ and finally gets comparable or even
exceeds all $\gamma_{M}$: the second and the third terms in (\ref{26a9}) are of the same order.
That is why we neglect the linear term in $\gamma_{M}$, but keep it in $\gamma_{a}$.

 In order to find the relationship between the primordial perturbation depth $\vartheta_1$ and
the present-day void underdensity we apply the method used in \citep{24, 25}. The time passed
between $a=a_1$ and $a=a_0$ is $\int^{a_0}_{a_1}\!\! \frac{da}{a H(a)}$, and similarly the time
passed between $b=b_1$ and $b=b_0$ is $\int^{b_0}_{b_1}\!\! \frac{db}{b \mathcal{H}(b)}$.
Apparently, the time intervals should be equal. Moreover, we may substitute the lower limit of
integration in both integrals by $0$: both integrals are very small and almost equal, if one takes
the integrals from $0$ to $a_1$ or $b_1$. Thus, we obtain:
\begin{eqnarray}
\int^{a_0}_{0}\!\! \frac{da}{a H(a)}=\int^{b_0}_{0}\!\! \frac{db}{b \mathcal{H}(b)}. \label{26a10}
\end{eqnarray}
The method based on the equating of ages of different parts of the Universe justifies our neglect
of radiation. Indeed, the present-day radiation density $\rho_{\gamma,0}< 10^{-4} \rho_{c,0}$
\citep{gorbrub1}. Radiation dominates in the early Universe, but this period is in time ($\sim
10^6$ years, which is very short with respect to the Universe age $\sim 14$~bln. years).

\begin{figure}
    \resizebox{0.9\hsize}{!}{\includegraphics{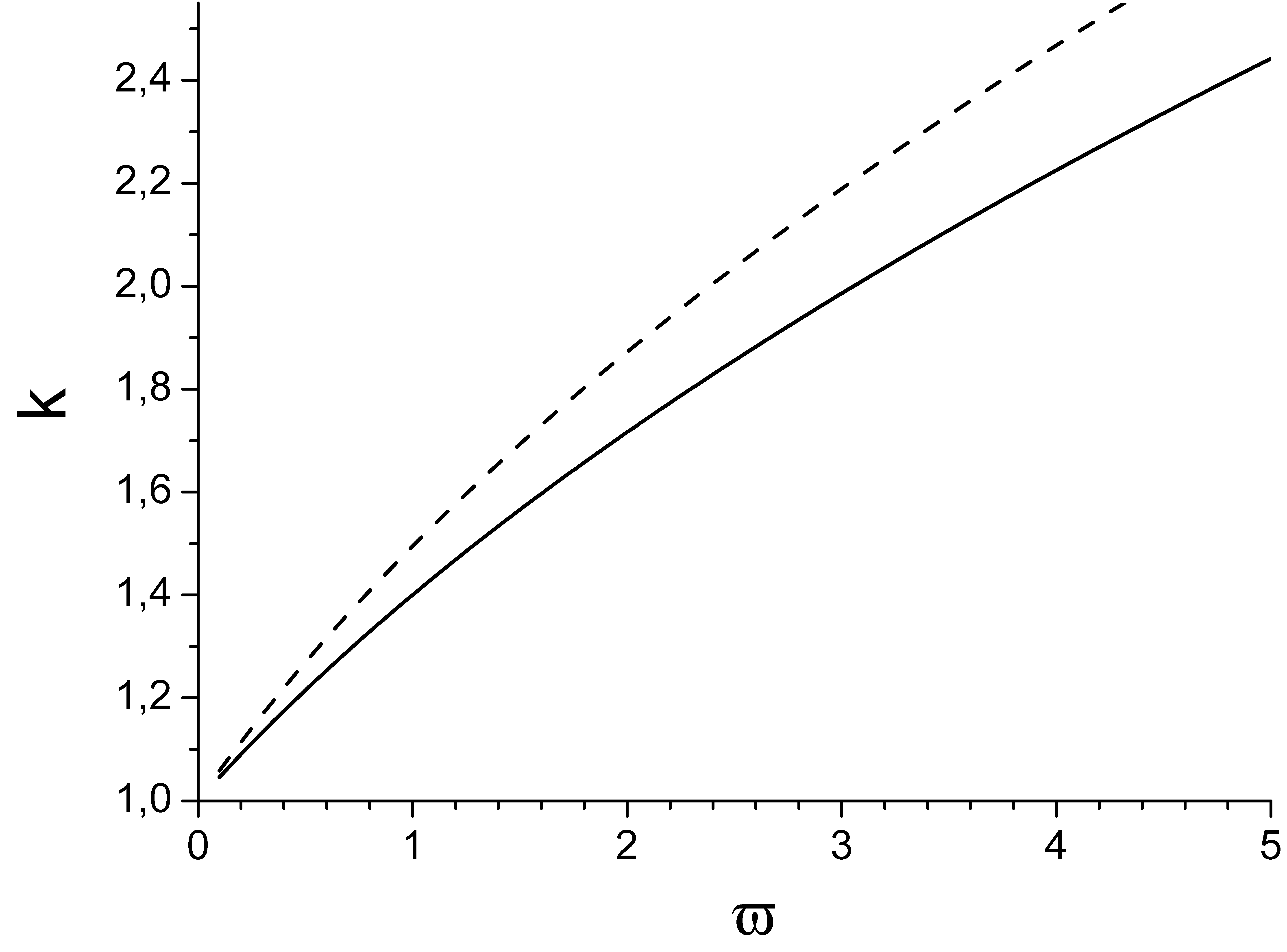}}
    \caption{The 'magnification ratio' $k$ at the void center as a function of $\varpi$ (see equation (\ref{26a13})).
We assume $\Omega_{M,0}=0.306$ (solid line). The case of $\Omega_{\Lambda}=0$, $\Omega_{M,0}=1$
(dashed line) is shown for comparison.} \label{26fig1}
\end{figure}

The first integral in~(\ref{26a10}) is equal to $t_0$ (see equation~(\ref{26a3})). We may rewrite
the second integral in~(\ref{26a10}), substituting there equation~(\ref{26a9}), and after simple
transformations we obtain:
\begin{equation}
\int^{b_0}_{0}\!\! \dfrac{db}{b \mathcal{H}(b)}=\int^{(b_0/a_0)}_{0}\!\!\!\!\!\!
\dfrac{\sqrt{x}dx}{H_0 \sqrt{x^3+\Omega_{M,0}(1+3\varpi x-x^3)}}.
 \label{26a11}
\end{equation}
Here we introduce a new notation $x=b/a_0$. We designate the 'magnification ratio' of the void
center by $k\equiv b_0/a_0$. Indeed, the central area of the void contains less matter than the
Universe in average and expands faster. As a result, the same amount of matter occupies larger
volume in the void center, and the present-day central density $\varrho$ of the void is
$\varrho=\rho_{c,0} \Omega_{M,0}/k^3$. The central void underdensity
\begin{equation}
\delta_c\equiv\frac{\varrho}{\rho_{M,0}}-1=\frac{\varrho}{\rho_{c,0}
\Omega_{M,0}}-1=\frac{1}{k^3}-1
 \label{26a12}
\end{equation}
is frequently used in literature. We should underline that $\varrho$ is the void central density,
averaged over smaller substructures (like separate galaxies). Therefore, though $\varrho$ is the
lowest matter density in the void if we consider its smoothed density profile, the density in the
void center may vary around this average value, following the small-scale structure inside the
void. Now we substitute equations~(\ref{26a3}) and~(\ref{26a11}) to~(\ref{26a10}):
\begin{equation}
\frac{\arcosh(\Omega^{-0.5}_{M,0})}{\sqrt{1-\Omega_{M,0}}}=\frac32\int\limits^{k}_{0}\!\!\!\!
\dfrac{\sqrt{x}dx}{ \sqrt{x^3+\Omega_{M,0}(1+3\varpi x-x^3)}}.
 \label{26a13}
\end{equation}
This equation defines $k$ as an implicit function of $\varpi$ and $\Omega_{M,0}$. We should
underline that deducting it we did not suppose that the final structure is linear or small, and it
is valid when the underdensity becomes nonlinear as well. The dependence is represented in
Figure~\ref{26fig1}.

We have obtained the solution assuming that $R_1\ll\lambda$. In order to estimate the radius of its
feasibility, we need to derive equation~(\ref{26a9}) substituting expansions~(\ref{26c2}) with
their quadratic terms. The quadratic corrections depend on the initial underdensity shape. In the
spherically-symmetric case equation~(\ref{26a13}) stays true, but $\varpi$ should be substituted by
$\varpi^*\equiv \varpi (1-38\pi^2 R^2_1/15\lambda^2_1)$. One may see that the quadratic correction
is only $17\%$ at $R_1=\lambda_1/12$, and so the Tolman approximation that we use is valid on the
third part of the initial underdensity size $\sim \lambda_1/2$.

\begin{figure}
    \resizebox{0.9\hsize}{!}{\includegraphics{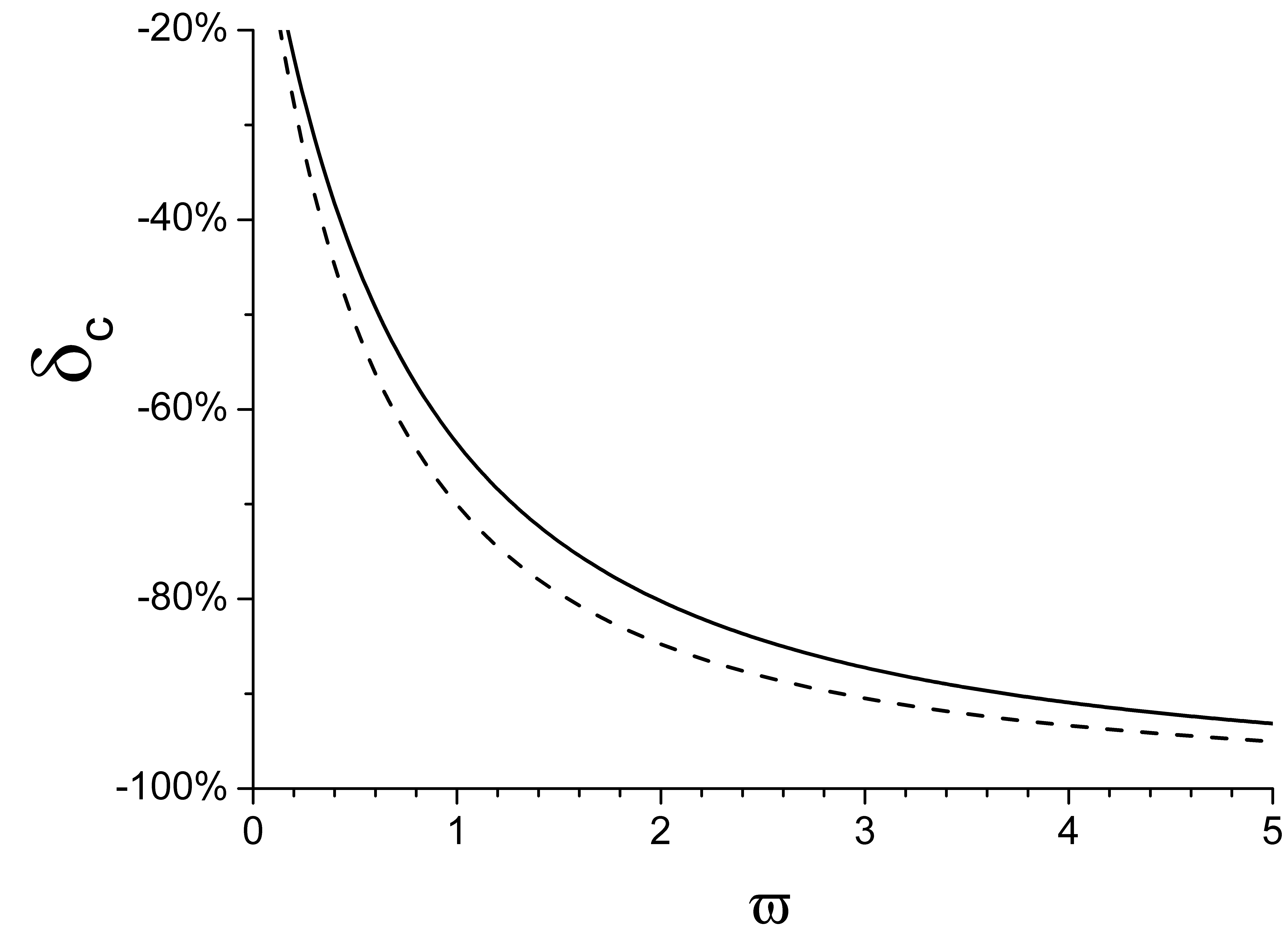}}
    \caption{The underdensity $\delta_c$ of a formed void at $z=0$,
as a function of amplitude parameter $\varpi$ of the initial perturbation. We assume
$\Omega_{M,0}=0.306$ (solid line). The case of $\Omega_{\Lambda}=0$, $\Omega_{M,0}=1$ (dashed line)
is shown for comparison.} \label{26fig2}
\end{figure}

We neglect tidal perturbations, and there are sufficient arguments to use this supposition. First
of all, voids have roundish, often almost spherical shape, which would be corrupted by significant
tidal effects. The void is surrounded by huge masses, but the honeycomb wall is rather
spherically-symmetric with respect to the void center. The matter outside the host honeycomb of the
void is also distributed quite spherically-symmetric because of the uniformity and isotropy of the
Universe. A spherically-symmetric matter distribution around a sphere does not create any
gravitational field inside the sphere, which is true in the general theory of relativity
\citep{zn2}, as well as in the Newtonian theory.

\section{Discussion}
\label{discussion26}

Equations~(\ref{26a12}) and~(\ref{26a13}) show that there is a one-to-one correspondence between
the present-day underdensity in the void center and the initial perturbation amplitude $\varpi$.
Cosmological observations suggest that the primordial perturbations are gaussian: if we randomly
choose a sphere of radius $R/h$ (where $h=H_0/100$~{km/(s$\cdot$ Mpc)}) at a moment when the
perturbations are still linear, the probability that the density contrast inside $R/h$ is equal to
$\vartheta$ is $(\sigma_R \sqrt{2\pi})^{-1} \exp(-\vartheta^2/2\sigma^2_R)$. The time dependence of
$\sigma_R$ looks like $\sigma_{R, t}\propto g(z)/(z+1)$, where multiplier $g(z)$ takes into account
the dark energy influence (see equation~(\ref{26b2}), $g(z=0)\equiv g_0= 0.78$). Cosmological
observations show that the present-day amplitude of the linear power spectrum on the scale of
$R_0=8/h\simeq 11$~{Mpc}  is equal to $\sigma_8=0.815\pm 0.009$ \citep{pdg18}. This scale
approximately corresponds to small voids. Comparing the equations of this paragraph, we find the
distribution for objects of $R_0=8 h^{-1}$~{Mpc} over parameter $\varpi$:
\begin{equation}
p(\varpi)=\frac{g_0}{\sigma_8 \sqrt{2\pi}}\exp\left(-\frac{\varpi^2 g^2_0}{2 \sigma^2_8}\right).
 \label{26b4}
\end{equation}
The regions with $0<\varpi\lesssim 0.5$ correspond to $|\delta_c|<45\%$ (see figure~\ref{26fig2})
and hardly can be considered as voids. Thus, the normalization factor in (\ref{26b4}) should be
corrected. Hereafter we accept this debatable void criterion ($\varpi\ge 0.5$, i.e.,
$\delta_c<-45\%$). Even then approximately a half of the voids of $R_0=8 h^{-1}$~{Mpc} have
$\varpi< 1.45$, i.e., $\delta_c > -73\%$, there are almost no voids ($\sim 1.6\%$) with $\varpi>
3\sigma_8/g_0\simeq 3.1$ ($\delta_c\le -88\%$), and even the voids with $\varpi>
2\sigma_8/g_0\simeq 2.1$ ($\delta_c\le -81\%$) are rare ($<20\%$). The probability of existence of
a void with large $|\delta_c|$ falls very rapidly: only $\sim 10^{-8}$ of voids can have
($\delta_c< -95\%$). Since $R_0=8 h^{-1}$~{Mpc} corresponds to small voids and $\sigma_R$ slowly
decreases with $R$, larger voids should have similar, but slightly lower underdensities.

There are, at least, two ways to estimate $\delta_c$ with the help of astronomical observations.
One of them is based on the galaxy number counts. The number density of galaxies inside voids is by
a factor of tens lower than the average one. However, the galaxy number depends only on the baryon
matter distribution, which is influenced by the baryon acoustic oscillations, contrary to the DM
distribution. Besides, typically we can see only relatively bright galaxies inside the void, and
their number is defined not only by the quantity of baryon matter, but also by the structure
formation details etc. Meanwhile, $\delta_c$ is on $\sim 80\%$ defined by the dark matter.

On the contrary, the second method of the observational determination of $\delta_c$ is based on the
gravitational lensing, being equally sensitive to the baryon and dark matter. Unfortunately, the
method is mainly sensitive to the fractional underdensity of the void as a whole (for instance,
\citet{voidobservation2015} found it to be equal to $\simeq -40\%$), and not to the central
underdensity $\delta_c$ that we have calculated. Figure~10 in \citep{fang2019} pictorially shows
that the observational data may be equally well fitted by models with small $\delta_c\simeq -40\%$
and almost flat density profile, as well as by models with extremely strong central underdensity
$\delta_c< -95\%$ and significant density growth towards the void walls. Thus, now $\delta_c$
cannot be confidently determined from astronomical observations.

Voids have been extensively modelled using N-body simulations. A popular model of the void density
profile is \citep[eqn. 15]{modelvoid}:
\begin{equation}
\frac{\rho(r)}{\rho_{M,0}}=A_0+A_3\left(\frac{r}{R_V}\right)^3.
 \label{26b5}
\end{equation} The authors
obtained the best fit $A_0=0.13\pm 0.01$, $A_3=0.70\pm 0.03$ for voids of $R_V=8 h^{-1}$~{Mpc}. The
fact that the void size in this fit coincides with that corresponding to $\sigma_8$ makes it
directly comparable with our results with $\varpi$ distributed as (\ref{26b4}). The value
$A_0=0.13$ corresponds to very deep central underdensity $\delta_c=A_0 - 1= -87\%$. Recent
high-resolution simulations result in the average underdensity of the whole void $\simeq -84\%$
\citep[table 2]{nbodyvoids2019}, which implies that $\delta_c$ can be as low as $\delta_c=A_0 - 1=
-90\%$ or even less. So N-body simulations suggest very low matter content of the voids.

Based on the obtained solution (defined by equations~(\ref{26a11}) and~(\ref{26a12})), we may make
several conclusions. First, our results suggest that N-body simulations somewhat overestimate the
emptiness of voids. Indeed, the central underdensity obtained by the simulations $\delta_c\sim
-85\%-90\%$ corresponds to $\varpi\simeq 2.6-3.7$. It is equal to $2.5-3.5$ standard deviations in
equation~(\ref{26b4}). The underdensity of the majority of voids should be lower $|\delta_c| <
73\%$. Perhaps, the reason of the this discrepancy are the recently reported convergency issues of
N-body simulations \citep{vanderbosch2018, 17, 20, 21}. The analytical solution provides us a good
opportunity to check the simulations: the value of $\varpi$ of the initial underdensity can be
easily calculated, and then one may compare $\delta_c$ of the void formed from it with the
analytical solution.

We can illustrate the low probability of $|\delta_c| \ge 85\%$ with the help of the above-mentioned
symmetry: an overdensity and an underdensity of the same size and absolute density contrast
$|\vartheta|\equiv |(\bar \rho - \rho)|/\bar \rho$ occur in the primordial perturbations with the
same probability. If we have a void with the central underdensity $\delta_c$, we may determine its
amplitude on the linear stage $\varpi$ (Figure~\ref{26fig2}), and then calculate the redshift
$z_{form}$ when a structure forms from a primordial overdensity of the same size and relative
amplitude $\varpi$ (with the help of (\ref{26b2})). We may see that $\delta_c=-85\%, -90\%, -95\%$
correspond to $z_{form}=1.53, 2.66, 5.37$, respectively. Even small voids of $R_0=8/h\simeq
11$~{Mpc} are formed from very extensive waves and correspond to overdensities of mass $\sim 2\cdot
10^{14} M_\odot$, i.e., to galaxy clusters. The cluster formation at $z=1.53$ is unlikely but
possible, while $z_{form}=2.66$ and especially $5.37$ seem to be too large.

As we have shown, the central region of a void can be considered as a part of an open Friedmann's
'universe'. It is important that its properties differ drastically from those of our Universe. We
may see from~(\ref{26a2}) that the Universe is matter-dominated until $z\simeq 0.314$, and then the
dark energy prevails. Equation~(\ref{26a9}) shows that the evolution of the 'universe' inside the
void center is more complex: the matter-dominated phase lasts only until $a_0/b=3\varpi$, i.e.,
until $z\simeq 2.58$ if $\varpi=1$ ($\delta_c=-64\%$), $z\simeq 6.0$ if $\varpi=2$
($\delta_c=-80\%$), and $z\simeq 9.4$ if the void is very deep ($\varpi=3, \delta_c=-87\%$). Since
then the curvature term prevails (this stage is absent in the Universe at all). Much later, at
$a_0/b=\sqrt{\Omega_{\Lambda,0}/3\varpi\Omega_{M,0}}$, the dark energy starts to dominate, but if
the void is very deep ($\varpi>2.925, |\delta_c|>87\%$), it has not happened yet, and these voids
are still dominated by the curvature term. Moreover, even the voids of the same size cannot have
the same $\delta_c$: $\delta_c$ and $\varpi$ are bound by one-to-one relation~(\ref{26a13}), and
$\varpi$ has Gaussian distribution~(\ref{26b4}).

The presence of the long and early starting stage of the curvature term prevalence profoundly alter
the structure formation in voids. A universe dominated by the curvature term expands significantly
faster than the matter-dominated one, and the perturbation growth is then significantly suppressed.
Even if $\varpi=1$ ($\delta_c=-64\%$) the curvature contribution is already remarkable during the
formation of gigantic galaxies ($z=3-4$). The perturbation suppression is stronger for larger
structures (they form later) and for voids with lower central density. If $\varpi=3$
($\delta_c=-87\%$), the curvature term prevails from $z=9.4$ till $z=0$, i.e., almost all galaxy
formation is affected by the curvature. The curvature suppression is the most probable reason why
the voids do not contain galaxy clusters and bright galaxies: these objects are the most massive,
develop late, and the faster expansion of the void just do not let them form.

We could see that the Tolman's approximation is applicable, at least, until $R_1=\lambda_1/12$ at
$z=z_1$. The density contrast in this toy 'universe' is $(\Delta\rho_M/\rho_M)_1 \approx
\vartheta_1 2\pi^2 R^2_1/\lambda^2_1\simeq \vartheta_1/7$ (see eqn.~(\ref{26c1})). During the
matter-dominated stage $(\Delta\rho_M/\rho_M)\propto b(t)$; when the curvature or dark energy
dominate, the perturbations grow slower. Thus, the the density contrast can grow no more than
$b_0/b_1=(z_1+1) k$ times. Since $\vartheta_1 (z_1+1)=\varpi$, we obtain the upper bound on the
density contrast in the present-day toy 'universe' $(\Delta\rho_M/\rho_M)_0 < k \varpi/7$.
Figure~\ref{26fig1} shows that typically $k\le 2$, and thus $(\Delta\rho_M/\rho_M)_0 <
\varpi/3.5<1$. The present-day matter density of the 'universe' is the central density of the void
$\varrho=\rho_{M,0}(1-\delta_c)\ll \rho_{M,0}$, and yet $\Delta\rho_M < \varrho$. First, it
justifies the use of the Tolman's solution. Second, it means that the density profile of the void
center is very flat: at $z=z_1$ the 'universe' occupies approximately a third of the underdensity
radius $\lambda_1/4$, but then this region expands $k$ times stronger than the Universe on average
and may occupy a half of the underdensity radius or even more. In principle, this result coincides
with phenomenological equation~(\ref{26b5}), which suggests $\Delta\rho_M/\varrho = 0.7$ at
$r=R_V/2$. However, equation~(\ref{26b5}) realizes the profile flatness by assuming that
$\rho_M\propto r^3$ in the center, while our analysis suggests that the central profile is still
quadratic, but with very small coefficient. Probably, a single power-law is not sufficient to fit
the full void profile, and it favors either more complex void models like \citep{hamaus2014} or two
separate models for the void bottom and the void walls.

Our consideration illustrates the fallacy of a widely accepted belief in necessity of the dark
energy for void formation. Equation~(\ref{26a6}) shows that $\gamma_{a,1}$ is entirely independent
of $\rho_{\Lambda}$, i.e., $\gamma_{a,1}$ is the same even if $\Lambda=0$, the underdensity expands
faster, and the void forms as well, though the void expansion law and the final $\delta_c$ are
completely different in this case. Moreover, the early analytical calculations (for instance,
\citet{bertschinger1985, weygaert2004}) disregard the cosmological constant  and suggest larger
void density contrasts $|\delta_c|$ than we obtain, which means that the dark energy suppresses
void growth. As we could see, this effect can be characterized by the factor $g(z)$, if the
perturbation is linear. We may qualitatively illustrate the suppression in the nonlinear regime as
well. For instance, if there is no dark energy and $\Omega_{M}=1$, then $a(t)\propto t^{2/3}$,
$b(t)\propto t$ when $t\to\infty$, i.e. the central region of the void will always expand stronger
than the universe on average, and $|\delta_c|$ will grow forever. On the contrary, the expansion
rates of voids and of the Universe on average have already become almost equal in the real
Universe: they are both defined by the main cosmological component, the nonzero cosmological
constant. When $t$ tends to infinity, $\mathcal{H}^2(t)={H}^2(t)=8 \pi G \rho_\Lambda/3$, and
$\delta_c$ becomes constant. Thus, the dark energy hampers the grows of underdensities, as well as
of overdensities. One may see it in figures~\ref{26fig1} and~\ref{26fig2}, where the case of
$\Omega_{\Lambda}=0$, $\Omega_{M,0}=1$ is shown by the dashed line for comparison.

We would like to thank the Heisenberg-Landau Program, BLTP JINR, and the Academy of Finland
mobility grant 1341541, for the financial support of this work. This research is supported by the
Munich Institute for Astro- and Particle Physics (MIAPP) of the DFG cluster of excellence "Origin
and Structure of the Universe".

\label{lastpage}
\end{document}